\pdfoutput=1

\documentclass[11pt]{article}

\usepackage{ACL2023}

\usepackage{times}
\usepackage{latexsym}

\usepackage[T1]{fontenc}

\usepackage[utf8]{inputenc}

\usepackage{microtype}

\usepackage{inconsolata}

\usepackage{caption}
\usepackage{graphicx}
\usepackage{float} 
\usepackage{subcaption}
\usepackage{amsmath}
\usepackage{amssymb}
\usepackage{hyperref}
\usepackage{url}
\usepackage{hyperref}
\usepackage{booktabs}
\usepackage{enumitem}
\usepackage{multirow}

\newcommand{\M}{AlignSTS}

\title{\M: Speech-to-Singing Conversion via Cross-Modal Alignment}

\author{
Ruiqi Li$^1$, 
Rongjie Huang$^1$, 
Lichao Zhang$^1$, 
Jinglin Liu$^2$,  
Zhou Zhao$^1$\thanks{* Corresponding author}
\\
$^1$Zhejiang University \\
\texttt{\{ruiqili,rongjiehuang,zju\_zlc,zhaozhou\}@zju.edu.cn} \\
$^2$ByteDance \\
\texttt{liu.jinglin@bytedance.com}
}

\begin{document}

\maketitle

\renewcommand{\thefootnote}{\fnsymbol{footnote}}

\begin{abstract}

The speech-to-singing (STS) voice conversion task aims to generate singing samples corresponding to speech recordings while facing a major challenge: the alignment between the target (singing) pitch contour and the source (speech) content is difficult to learn in a text-free situation. This paper proposes \M, an STS model based on explicit cross-modal alignment, we 1) adopt a novel rhythm adaptor to predict the target rhythm representation to bridge the modality gap between content and pitch, where the rhythm representation is disentangled in a simple yet effective way and is quantized into a discrete space; and 2) leverage the cross-modal aligner to re-align the content features explicitly according to the predicted rhythm and conduct a cross-modal fusion for re-synthesis. Experimental results show that \M~achieves superior performance in terms of both objective and subjective metrics. Audio samples are available at \href{https://alignsts.github.io}{https://alignsts.github.io}.


\end{abstract}

\section{Introduction}

Speech-to-singing (STS) voice conversion \cite{saitou2004analysis,cen2012template,parekh2020speech} aims to transfer speech samples into the corresponding singing samples, with the timbre identity and phoneme information unaltered. An STS system takes a speech sample and a target melody as conditions and then generates a high-quality singing sample following the target musical melody. 
Speech-to-singing research is important for human voice study and useful for practical applications such as computer-aided music production or musical entertainment.

Researchers have developed three major STS approaches: 1) Model-based approaches\citep{saitou2004analysis,saitou2007speech} use phone-score synchronization information to manually align the phonemes and the target musical notes with artificial control models. 2) Template-based STS~\citep{cen2012template,vijayan2017dual,vijayan2018analysis} requires an available high-quality reference vocal input, which will be aligned with the input speech for subsequent musical feature extractions. The alignment is the key part and is mostly based on dynamic time warping (DTW). 3) Style transfer approach~\citep{parekh2020speech} views STS as a style-transfer problem. This class of methods considers the specific properties that are transformed during conversion as "style". 
\citet{parekh2020speech} stretch the input speech to the same length as the target F0 contour and concatenate the latent features to fuse the asynchronous representations. \citet{wu2020speech} serve as a continuation and extension of their prior work by leveraging boundary-equilibrium GAN (BEGAN) ~\citep{berthelot2017began}. 



Despite their recent success, however, the style information is complex and is composed of multiple entangled features in the time domain (the duration information and the temporal length) and the frequency domain (the pitch information). Simply stretching the representations temporally or applying implicit self-attention can cause alignment problems.
In a larger sense, human voice information (such as speech~\citep{huang2023audiogpt} and singing~\citep{huang2021multi,huang2022singgan}) is composed of several variance information, each controlling a specific sensory modality. 
Plenty of arts focus on the decomposition and resynthesis of speech signal \cite{qian2020unsupervised,chan2022speechsplit2,choi2021neural,choi2022nansy++,huang2022fastdiff,huanggenerspeech}. They managed to roughly decompose speech signals into components like linguistic content, pitch, rhythm, timbre identity, etc. By manipulating any of these components, one can resynthesize a customized speech waveform. 
In the context of STS, the components that will be manipulated are 1) the frequency modality, namely the pitch information; and 2) the duration modality, or rhythm. The challenge remains, however, since the rhythm information can be indeterminate given only the content and the target pitch information.

In this paper, we present a novel approach \M~ based on modality disentanglement and cross-modal alignment.
To tackle the alignment problem, we introduce a novel rhythm representation to bridge the modality gap between content and pitch. A rhythm adaptor is designed to predict the target rhythm representation, which is used to guide the content features to perform the alignment and the cross-modal fusion. 
Further, we explore speech-to-sing conversion in zero-shot scenarios, where we train the model with singing data in a self-supervised manner and test it on unseen speech data.
We categorize \M~as one of the style transfer approaches since we only need a source speech and a target pitch contour for conversion.
Experimental results demonstrate significant performance improvements over baseline models. The main contributions of this work are summarized as follows:
\begin{itemize}[leftmargin=*]
    \setlength{\itemsep}{0pt}
    \item We leverage the temporal modality, rhythm, to bridge the modality gap between the speech content and the target pitch. The rhythm representation is carefully designed and quantized into discrete space to model temporal dynamic states.
    \item We propose \M, an STS model based on cross-modal alignment, which predicts the target rhythm representation and uses it to conduct an explicit cross-modal alignment to re-align the content information. 
    \item Experimental results demonstrate that \M~achieves state-of-the-art results in both objective and subjective evaluations. Over baseline models, \M~reaches an absolute improvement of 0.39 in MOS for overall quality and 0.36 for prosody naturalness.
    
\end{itemize}

\section{Related Works}

\subsection{Voice Conversion}

Voice conversion (VC) focuses on changing the timbre identity of an utterance to a target speaker while keeping the content information (phoneme sequence intact). Inspired by image style transfer, CycleGAN-VC \cite{kaneko2018cyclegan} combines CycleGAN \cite{zhu2017unpaired} with gated CNN and identity-mapping loss to capture sequential and hierarchical structures. Similarly, StarGAN-VC \cite{kameoka2018stargan} adapts StarGAN \cite{choi2018stargan} and manages to train without parallel utterances and pay more attention to real-time processing. Apart from GANs, conditional variational autoencoders (CVAE) are also an important class of approaches. VAE-VC \cite{hsu2016voice} uses the encoder of a VAE to learn the speaker-independent phoneme representations, thus disentangling the timbre identity. ACVAE-VC \cite{kameoka2019acvae} notices that VAEs easily ignore the attribute class label input, i.e. the speaker identity, which therefore utilizes an auxiliary speaker classifier. AutoVC \cite{qian2019autovc} carefully designs a bottleneck mechanism within a simple autoencoder to achieve zero-shot many-to-many voice conversion with non-parallel data. 

\subsection{Speech Representation Disentanglement}

Human speech~\citep{huang2023make,huang2022transpeech} is a severely complicated and comprehensive information stream, where a number of latent units entangled with each other such as content, pitch, timbre, etc. The disentanglement of the speech signal is an attempt to learn factorized and even interpretable generative factors \cite{bengio2013representation} for further application, like style transfer or domain adaptation. 
NANSY \cite{choi2021neural} manipulates disentangled factors such as content, pitch, and speed of speech signal, where content and pitch are decomposed by wav2vec 2.0 \cite{baevski2020wav2vec} and Yingram, respectively. Information bottleneck \cite{qian2019autovc} is also a popular way to disentanglement. 
Following AutoVC, SpeechSplit \cite{qian2020unsupervised} introduces three carefully designed information bottlenecks to improve decomposition. VoiceMixer \cite{lee2021voicemixer} leverages a similarity-based information bottleneck and adversarial feedback to disentangle content and voice style. SpeechSplit 2.0 \cite{chan2022speechsplit2} alleviates the bottleneck tuning in SpeechSplit by applying efficient signal processing techniques on encoder inputs. 


\begin{figure*}[htbp]
\centering
\includegraphics[width=\textwidth]{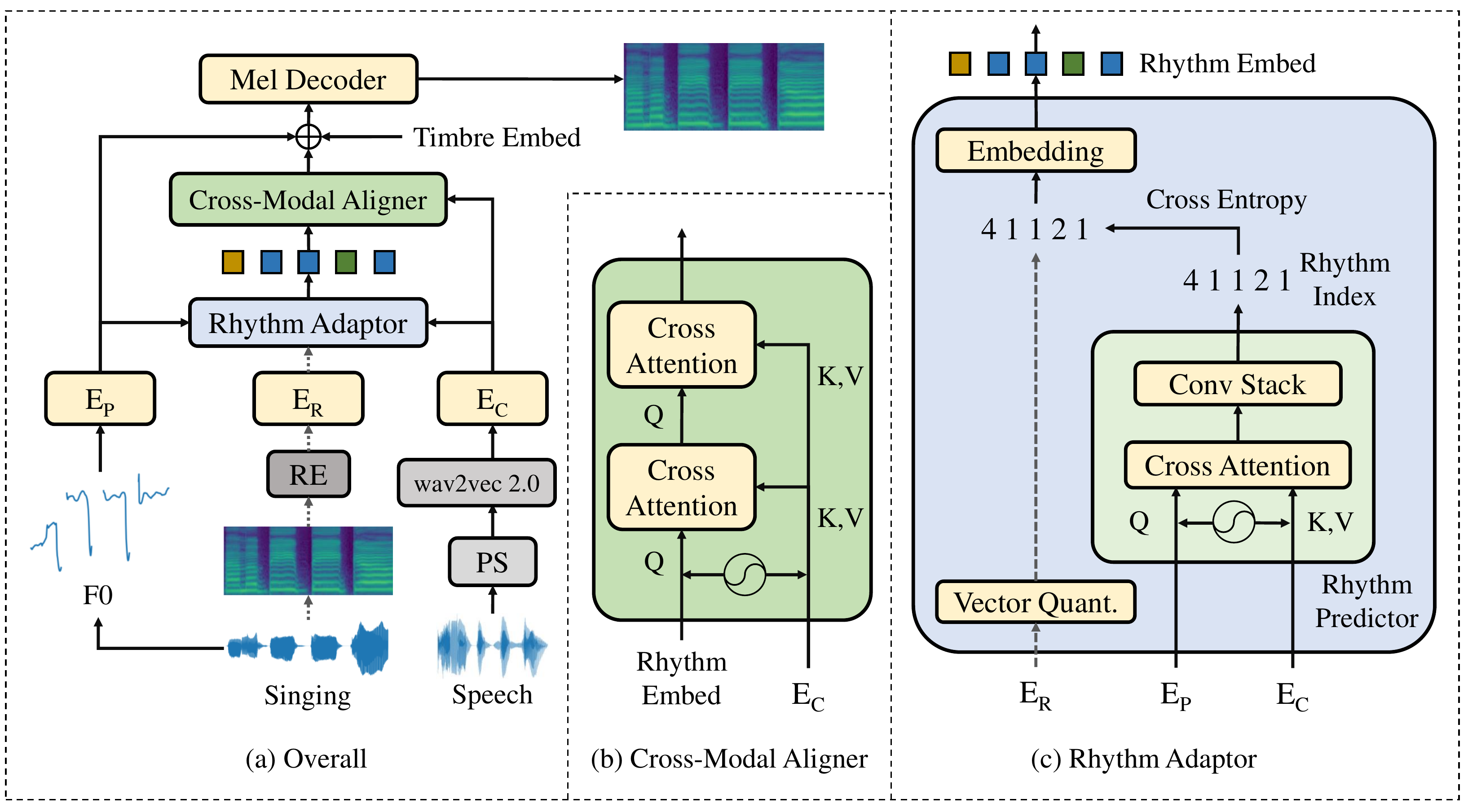}
\setlength{\abovecaptionskip}{-0.4cm}
\setlength{\belowcaptionskip}{-0.4cm}
\caption{The architecture of the proposed model. Subfigure (a) is the overall structure. "PS" stands for pitch smooth operation. "RE" denotes extracting rhythm information. "E$_\text{P}$", "E$_\text{R}$", and "E$_\text{C}$" stand for three encoders used to encode pitch, rhythm, and content information, respectively. The parts connected by gray arrows on the dotted lines are only involved in the training stage and are deactivated during inference. In subfigures (b) and (c), "E$_\text{P}$", "E$_\text{R}$", and "E$_\text{C}$" denote the output from each encoder. In subfigure (c), a CE loss is applied to the rhythm indices predicted by the rhythm predictor during the training stage. }
\label{fig:main-model}
\end{figure*}

\section{\M}

In this section, we first define and formulate the problem of speech-to-singing (STS) voice conversion. We then present the information perturbation methods, the rhythm modality modeling, and the cross-modal fusion mechanism. Finally, we introduce the overall architecture of \M~and the training/inference procedure.

\subsection{Problem Formulation}


Let $S_{\text{sp}}$ and $S_{sg}$ denote the spectrograms extracted from speech and singing signals. We assume that vocal signals are a comprehensive fusion of several variance information, i.e., content, pitch, rhythm, and timbre. Therefore, this process can be formally defined in \autoref{eq:formulation1},
where $\boldsymbol{c}_{\text{sp}}$, $\boldsymbol{i}_{\text{sp}}$, $\boldsymbol{f}_{\text{sp}}$, and $\boldsymbol{r}_{\text{sp}}$ denote the representations of content, timbre identity, pitch, and rhythm information of speech data. $g(\cdot)$ denotes the multi-modality fusion. 
\begin{align}
    S_{\text{sp}} &= g(\boldsymbol{c}_{\text{sp}}, \boldsymbol{i}_{\text{sp}}, \boldsymbol{f}_{\text{sp}}, \boldsymbol{r}_{\text{sp}}) \label{eq:formulation1} \\
    \widehat{S_{\text{sg}}} &= f_\theta(\boldsymbol{c}_{\text{sp}}, \boldsymbol{i}_{\text{sp}}, \boldsymbol{f}_{\text{sg}}, \boldsymbol{r}_{\text{sg}}(\boldsymbol{c}_{\text{sp}}, \boldsymbol{f}_{\text{sg}})) \label{eq:formulation2}
\end{align}
The problem of STS is to learn a neural network $f_\theta$, such that if we switch the pitch component $\boldsymbol{f}_{\text{sp}}$ to target pitch feature $\boldsymbol{f}_{\text{sg}}$, the corresponding singing spectrogram $\widehat{S_{\text{sg}}}$ will be generated
while preserving the content and timbre intact, as shown in \autoref{eq:formulation2}. Note that the rhythm modality implies temporal information, which will also be influenced and should be adapted according to the pitch and content. $\boldsymbol{r}_{\text{sg}}$ is the adapted rhythm representation and is generated conditioned on $\boldsymbol{f}_{\text{sg}}$ and $\boldsymbol{c}_{\text{sp}}$.

\subsection{Method Overview}
\M~treats speech and singing signals as a comprehensive fusion of several variance information, which can be further regarded as different sensory modalities.
Pitch and rhythm features are the main modalities to convert in STS, which can be disentangled in parallel. However, the synthesis logic of variance information needs to be carefully designed during conversion. A phoneme sequence and a pitch contour seem to be uncorrelated with each other at first glance, yet it is highly possible for a human to create a suitable alignment and produce a singing melody. The mechanism behind the human behavior is: (1) find an appropriate sequence of onset and offset timings of phonemes and notes, or as known as rhythm, according to the lyrics and the melody; and
(2) place the time-stretched phonemes in order according to the rhythm and combine them with the melody to produce the singing result.
Inspired by this mechanism, \M:
\begin{enumerate}[leftmargin=*]
    \setlength{\topsep}{0pt}
    \setlength{\itemsep}{-2pt}
    \item Decompose the input speech signal into several disentangled variance information.
    \item According to the altered speech component, i.e., the pitch contour, adapt the speech component that controls temporal duration information, namely the rhythm representation.
    \item Perform a cross-modal alignment to re-align the content according to the adapted rhythm representations and carry out a modality fusion to combine the variance information.
\end{enumerate}


\subsection{Information Perturbation}

Speech and singing samples are fully complex and can be decomposed into variance information such as content, pitch, rhythm, and timbre. Each feature needs to be extracted and disentangled from the other, which can be achieved by certain information perturbations.

\begin{itemize}[leftmargin=*]
    \setlength{\topsep}{0pt}
    \setlength{\itemsep}{0pt}
    \item \textbf{Linguistic Content.} We use a wav2vec 2.0 \cite{baevski2020wav2vec} model pre-trained and fine-tuned on 960 hours of Librispeech on 16kHz sampled speech audio \footnote{https://huggingface.co/facebook/wav2vec2-base-960h} to extract the linguistic content. It is shown that extracted features from speech SSP models such as wav2vec 2.0 can be applied to downstream tasks like ASR,
    indicating that the extracted features should provide rich isolated linguistic information. 
    Prior works like SpeechSplit \cite{qian2020unsupervised} and AutoPST \cite{pmlr-v139-qian21b} utilize random sampling to perturb the rhythm information within the content, given their input and output audios are the same speech. Using the paired speech and singing data, we do not need extra random sampling but leverage the natural discrepancy between speech and singing to perturb the rhythm within the content information.
    \item \textbf{Pitch.} We extract the fundamental frequency contour F0 of singing data as pitch information. 
    The fundamental frequency contour is then quantized to 256 possible values uniformly. 
    The F0 contour contains minimum rhythm information, given a common singing situation where one phoneme corresponds to several musical notes (or vice versa), making the rhythm theoretically indeterminate.

    \item \textbf{Rhythm.} Rhythm is a crucial speech component that controls the overall speed, the duration of each phoneme, and the pattern of the onset and offset of syllables. Therefore, the rhythm modality provides the duration information for both content and pitch. A good rhythm representation creates a "fill in the blank" mechanism \cite{qian2020unsupervised} for content information to re-align. 
    Meanwhile, the patterns of the rhythms of speech and singing voices differ significantly, the intensity of singing voices is generally more fluctuating and distinct than speech.
    Therefore, we directly utilize the time-domain energy contour $\boldsymbol{e}_t$ of singing samples, which is computed by taking the L2-norm of all the frequencies for each time step.
    To eliminate the relative fluctuation and leave only the rhythm information, we further normalize the energy contour $\boldsymbol{e}_t$ using the Sigmoid function $\sigma(\cdot)$:
        \begin{equation} \label{eq:rhythm}
            \boldsymbol{r}_t = \sigma \left(\beta \times \frac{\boldsymbol{e}_t - \text{mean}(\boldsymbol{e})}{\text{std}(\boldsymbol{e}) + \epsilon}\right)
        \end{equation}
    where $\epsilon$ is used to avoid division-by-zero error and $\beta$ is a hyperparameter used to control the normalizing effect. $\boldsymbol{r}_t$ is the resulting rhythm representation. We further stabilize the representation by applying Gaussian filtering with a standard deviation $\sigma$ of $1.0$. This scalar representation disentangles the rhythm information the most thoroughly while preserving the duration information. A visualization is shown in \autoref{fig:rhythm}.
        \begin{figure}[htbp]
            \vspace{-0.4cm}     
            \centering
            \includegraphics[width=0.48\textwidth]{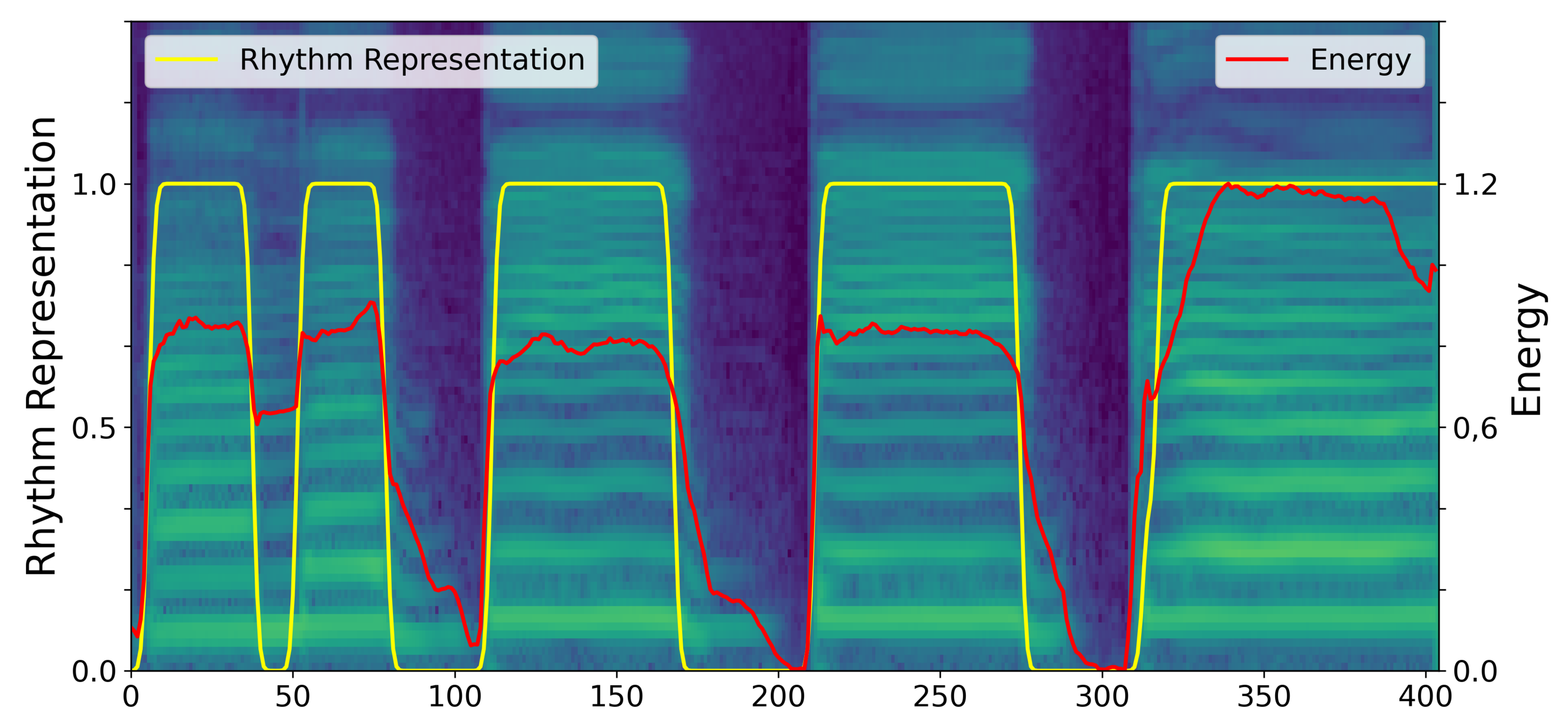}
            \setlength{\abovecaptionskip}{-0.2cm}
            \setlength{\belowcaptionskip}{-0.4cm}
            \caption{A visualization of rhythm representation with $\beta = 400$. The mel-spectrogram is extracted from a singing utterance "I'm a big big girl". A clear duration pattern is revealed and the relative intensity difference across phonemes is eliminated.}
            \label{fig:rhythm}
        \end{figure}
    \item \textbf{Timbre.} We leverage an open-source speaker identity encoding API Resemblyzer \footnote{https://github.com/resemble-ai/Resemblyzer} to extract the timbre representations. Resemblyzer generates a 256-dimensional vector that summarizes the characteristics of the given voice spoken. 
\end{itemize}

\subsection{Rhythm Modality Modeling}

The rhythm modality of the singing signal can be considered as a series of discrete temporal dynamic states, such as attack, sustain, transition, silence, etc. In automatic music transcription (AMT), the recognition of states refers to onset/offset detection.\cite{chang2014pairwise,fu2019hierarchical}
Similarly, traditional ASR methods like Kaldi \cite{Povey_ASRU2011} model the intra- and inter-states of phonemes to improve distinguishability. 
From this perspective, the rhythm modality is a "softened" version of duration with more intermediate states. 
The onset and offset states (intermediate states) of singing may last longer than that of speech (consider the fade-out effect). 
Inspired by this, we adopt a \textbf{Vector Quantization} (VQ) module to quantize the continuous rhythm features to model these temporal states and form an information bottleneck \cite{van2017neural}. The designed discrete latent embedding space can be denoted as $e \in \mathbb{R}^{K \times D}$ where $K$ is the number of clustered categories and $D$ is the dimension of each latent embedding vector $e_k \in \mathbb{R}^D, k \in 1,2,...,K$. A commitment loss \cite{van2017neural} is used to constrain the input representation to commit to a discrete embedding:
\begin{equation}
    \mathcal{L}_{\text{C}} = \left\|z_e(x) - \text{sg}[e]\right\|_2^2
\end{equation}
where $z_e(\cdot)$ denotes the VQ module and $\text{sg}$ denotes the stop-gradient operator.

To model the target rhythm representations conditioned on both input content information and target pitch contour, we design a \textbf{Cross-Attention} module by adopting the Scaled Dot-Product Attention \cite{vaswani2017attention}. 
The encoded pitch representation $X_P$ is used as the query, and the encoded content representation $X_C$ is used as both the key and the value. A positional encoding embedding is added to both representations before the attention module. The attention mechanism can be formulated as:
\begin{align}   \nonumber
    \text{Attention}&(Q, K, V)  \\ \nonumber
                    &= \text{Attention}(X_P, X_C, X_C) \\ 
                    &= \text{Softmax}\left(\frac{X_P X_C^T}{\sqrt{d}}\right) X_C
\end{align}
where $X_P$ and $X_C$ are first projected to the query, the key, and the value representations in practice. 

Since the target rhythm features are not available during inference, we only activate the VQ module and use the discrete embeddings generated from it for downstream modules during the training stage. The cross-attention module and a stack of convolution layers are combined to serve as the \textbf{Rhythm Predictor}, 
which takes the generated discrete rhythm embeddings from the VQ module as the training target.
A cross-entropy (CE) loss is applied to train the rhythm predictor so it can generate the desired rhythm embeddings during inference:
\begin{equation}
    \mathcal{L}_{\text{R}} = - \frac{1}{T}\sum_{t=1}^T\sum_{c=1}^K y_{t,c} \log(\hat{y}_{t,c})
\end{equation}
where $T$ denotes the number of time frames. $y_{t,c}$ denotes the rhythm embedding indices generated from the VQ module, where $y_{t,c} = 1$ if $c = k_t$ and $k_t$ is the rhythm index at time step $t$. $\hat{y}_{t,c}$ denotes the predicted rhythm indices. 

\subsection{Cross-Modal Alignment}


With the target rhythm sequence generated, we design a cross-modal aligner to place the linguistic features along the time axis according to the target rhythm. This aligner uses rhythm information to bridge the gap between content and pitch modalities. We simply use a stack of two cross-attention layers mentioned before to complete this task. The rhythm embedding $X_R$ generated from the rhythm adaptor is used as the query, and the encoded content representation $X_C$ is again used as both the key and value. 

The alignment using the cross-attention mechanism can be considered a soft-selection operation over the linguistic content representations $X_C$ along the time axis at each target time step. Ideally, the resulting attention weight matrix should show a monotonic pattern and the alignment path should be concentrated and nearly diagonal. This requires extra constraints and regulations to make sure the model does not bypass the attention mechanism and simply interpolate the input content to the same length as the required rhythm representation. We mainly apply two techniques: windowing \cite{chorowski2015attention} and guided attention loss \cite{tachibana2018efficiently}, which is described in detail in \autoref{appendix:attention}.

With each variance information re-aligned, the content, the rhythm, and the pitch representations should have the same temporal length. We apply the inter-modality fusion across these features by element-wise vector addition. Furthermore, we involve the timbre information by adding the timbre embeddings extracted earlier. 

\subsection{Architecture}

\M~mainly consists of four modules: the encoders, the rhythm adaptor, the cross-modal aligner, and the mel decoder. The overall architecture of \M~is presented in \autoref{fig:main-model}. The detailed description and the overall hyperparameter setting are listed in \autoref{appendix:architecture}.


To accelerate the re-synthesis process and improve the audio quality, we adopt the teacher model of ProDiff \cite{huang2022prodiff}, a 4-step generator-based diffusion model, to be the mel decoder. A generator-based diffusion model parameterizes the denoising model by directly predicting the clean data, instead of estimating the gradient of data density. Therefore, the generator-based method bypasses the difficulty of predicting sample $\boldsymbol{x}_t$ using a single network at different diffusion steps $t$, which allows us to train the decoder with a reconstruction loss. We use two objectives to be the reconstruction loss:
\begin{itemize}[leftmargin=*]
    \setlength{\itemsep}{0pt}
    \item \textbf{Mean Absolute Error (MAE).} We apply MAE at each random term of the diffusion step $t$:
        \begin{align}
        \mathcal{L}_{\text{MAE}} &= \left\|\boldsymbol{x}_\theta\left(\alpha_t\boldsymbol{x}_0 + \sqrt{1 - \alpha_t^2}\boldsymbol{\epsilon}\right) - \boldsymbol{x}_0\right\|
        \end{align}
    where $\alpha_t$ denotes a derived constant that $\alpha_t = \prod\limits_{i=1}^t \sqrt{1 - \beta_i}$, in which $\beta_t$ is the predefined fixed noise schedule at diffusion step $t$.
    $\boldsymbol{\epsilon}$ is randomly sampled and $\boldsymbol{\epsilon} \in \mathcal{N}(0, \boldsymbol{I})$. $\boldsymbol{x}_0$ denotes the clean data and $\boldsymbol{x}_\theta$ denotes the denoised data sample predicted by the denoising neural networks $\theta$. 
    \item \textbf{Structural Similarity Index (SSIM) Loss.} We adopt SSIM \cite{wang2004image}, one of the state-of-the-art perceptual metrics to measure image quality, to tackle the problem of over-smoothness \cite{ren2022revisiting}:
    \begin{align}
        \mathcal{L}&_{\text{SSIM}} = 1 - \\ \nonumber
                   &\text{SSIM}\left(\boldsymbol{x}_\theta\left(\alpha_t\boldsymbol{x}_0 + \sqrt{1-\alpha_t^2}\boldsymbol{\epsilon}\right), \boldsymbol{x}_0\right)
    \end{align}
    where $\text{SSIM}(\cdot)$ is the SSIM function and is between 0 and 1. 
\end{itemize}

\subsection{Training and Inference}
The final loss of \M~consists of the following loss terms: 
1) the commitment loss of VQ module $\mathcal{L}_{\text{C}}$; 2) the CE loss of rhythm predictor $\mathcal{L}_{\text{R}}$; 3) the guided loss for cross-attention $\mathcal{L}_{\text{attn}}$; 4) the MAE reconstruction loss $\mathcal{L}_{\text{MAE}}$; and 5) the SSIM loss $\mathcal{L}_{\text{SSIM}}$. 
During the training stage, the cross-modal aligner takes the rhythm embeddings generated from the VQ module as input directly. At the same time, the rhythm predictor is trained to predict the correct indices of quantized vectors for rhythm embeddings using CE loss. During the inference stage, the VQ module is deactivated and the predicted rhythm indices from the rhythm predictor are used to look up the embeddings for subsequent modules. 

\section{Experiments}

\subsection{Experimental Setup}



\subsubsection{Dataset}

We utilize a subset of the PopBuTFy database \cite{liu2022learning} as our dataset. PopBuTFy is originally used for the singing voice beautifying (SVB) task, which consists of paired amateur and professional singing recordings. Additionally, we collected and annotated the speech version of a subset of PopBuTFy to create a paired speech and singing dataset. In all, the dataset consists of 152 English pop songs ($\sim$5.5 hours in total) and the respective speech recordings ($\sim$3.7 hours in total) from 16 singers. More details are listed in \autoref{appendix:dataset}.

\subsubsection{Implementation Details}


We use mel-spectrograms extracted from singing samples to be the training target. We transform the raw waveform with the sampling rate of 24000 Hz into mel-spectrograms with window size 512 and hop size 128. We extract the fundamental frequency contour $F_0$ using Parselmouth \cite{parselmouth,praat} as pitch information. In addition, we remove all the silent frames to accelerate the training process. The output mel-spectrograms are transformed into audio waveforms using a HiFi-GAN vocoder \cite{kong2020hifi} trained with singing data in advance. More details are listed in \autoref{appendix:architecture}.


\subsubsection{Training and Evaluation}

We train the proposed model on a single NVIDIA GeForce RTX 3090 GPU with a batch size of 20 sentences for 200k steps. 
The performance evaluation consists of two parts, objective and subjective evaluations, respectively.
\begin{itemize}[leftmargin=*]
    \setlength{\itemsep}{0pt}
    \item \textbf{Objective evaluation}. Following \cite{parekh2020speech}, we use log-spectral distance (LSD) and F0 raw chroma accuracy (RCA) using mir\_eval \cite{raffel2014mir_eval} as the objective metrics. LSD is computed by taking the average of the Euclidean distance between the predicted log-spectrogram and the ground truth recordings. For RCA, we set the maximum tolerance deviation as 50 cents. 
    In addition, we design a new evaluation metric, rhythm representation distance (RRD), to measure rhythm reconstruction performance by computing the Euclidean distance between the rhythm representations described in \autoref{eq:rhythm}. 
    
    \item \textbf{Subjective evaluation}. We conducted crowd-sourced mean opinion score (MOS) listening tests for subjective evaluation. Specifically, MOS-Q indicates the overall quality of the audio and MOS-P indicates the naturalness and coherence of prosody. The metrics are rated from 1 to 5 and reported with 95\% confidence intervals. 
\end{itemize}

\begin{figure*}[ht]
\centering
\includegraphics[width=\textwidth]{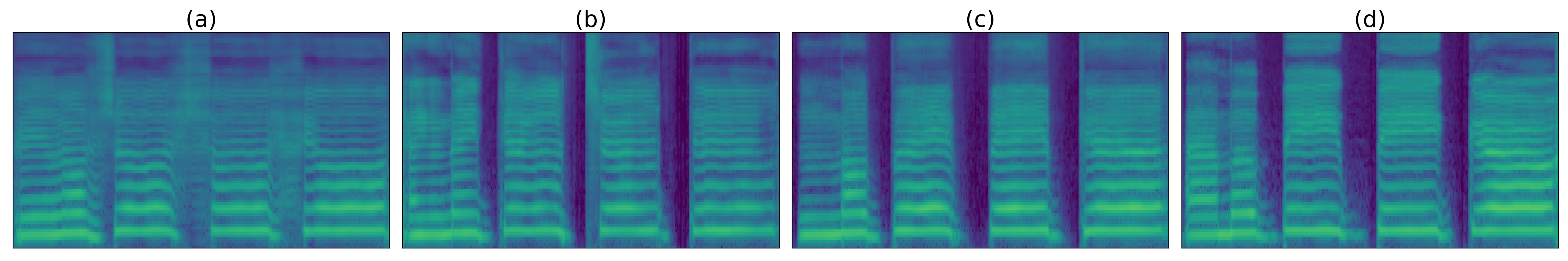}
\setlength{\abovecaptionskip}{-0.4cm}
\setlength{\belowcaptionskip}{-0.4cm}
\caption{Visualization of the output mel-spectrogram of each system. (a) \textit{\cite{parekh2020speech}}; (b) \textit{SpeechSplit 2.0 (w/o SE)}; (c) \textit{\M~(ours)}; (d) \textit{GT Mel}} .
\label{fig:mels}
\end{figure*}

\begin{table*}
\setlength\tabcolsep{10pt}
\centering
\setlength{\belowcaptionskip}{-0.4cm}
\begin{tabular*}{0.9\hsize}{l|ccc|cc}
\toprule
\bf Method                          & \bf LSD $\downarrow$  & \bf RCA $\uparrow$    & \bf RRD $\downarrow$  & \bf MOS-Q $\uparrow$  & \bf MOS-P $\uparrow$  \\
\midrule
\textit{GT Mel}                     & 2.8974                & 0.9959                & 0.3693                & 4.04$\pm$0.10         & 4.18$\pm$0.17         \\
\midrule[0.2pt]
\textit{\cite{parekh2020speech}}    & 7.3613                & 0.9218                & 0.7865                & 2.86$\pm$0.12         & 2.91$\pm$0.11         \\
\textit{SpeechSplit 2.0 (w/o SE)}   & 5.7681                & 0.9870                & 0.8262                & 3.19$\pm$0.06         & 3.45$\pm$0.13         \\
\midrule[0.2pt]
\textit{\M~(GAN)}                   & 5.4926                & 0.9875                & 0.5709                & 3.41$\pm$0.11         & 3.76$\pm$0.17         \\
\textit{\M~(ours)}                  & \bf 5.0129            & \bf 0.9934            & \bf 0.5366            & \bf 3.58$\pm$0.19     & \bf 3.81$\pm$0.09     \\
\midrule[0.2pt]
\textit{\M~(zero-shot)}             & 5.6607                & 0.9871                & 0.5693                & 3.29$\pm$0.08         & 3.46$\pm$0.11         \\
\bottomrule
\end{tabular*}
\caption{\label{tab:main-results}
The Objective and Subjective evaluation results of STS systems. 
}
\end{table*}

\subsubsection{Baseline Models}

We compare the quality of the generated audio samples of \M~with other approaches, including 
1) \textit{GT Mel}, in which we first convert the reference audio into mel-spectrograms and then convert them back to audio using HiFi-GAN; 
2) \cite{parekh2020speech}, an STS model based on encoder-decoder framework; 
3) \textit{SpeechSplit 2.0 (w/o SE)} \cite{chan2022speechsplit2}, where we train the model only conditioned on the target pitch contour, while the rhythm input is implemented by interpolating the source spectral envelope (from speech) to match the length of pitch contour (in both SpeechSplit 2.0 baselines, we interpolate the content input to match the target length); 
4) \textit{\M~(zero-shot)}, where we conduct zero-shot STS and test the model on unseen speech samples; and
5) \textit{\M~(GAN)}, where we change the diffusion mel decoder to the decoder of FastSpeech 2 \cite{ren2020fastspeech}, one of the SOTA approaches of non-autoregressive text-to-speech (NAR-TTS), combining a multi-window discriminator \cite{wu2020adversarially}. 
In addition, since SpeechSplit 2.0 is not originally designed for STS tasks, we conduct extra experiments on baseline \textit{SpeechSplit 2.0 (w/ SE)} in \autoref{appendix:extensional-experiments}, where we involve the target rhythm representations. 


\subsection{Main Results}

The results are shown in \autoref{tab:main-results}. The quality of \textit{GT Mel} is the upper limit of STS systems. In both objective and subjective evaluations, \M~outperforms the baseline systems by a large margin.
In both objective and subjective evaluations, the results of RCA and MOS-P are better than LSD and MOS-Q, which indicates that the condition of the target pitch contour possesses rich melody and prosody information, making the melody transfer much more effortless than phoneme modeling. The main challenge remains in the coherence and recognizability of phonemes. 
The results of RRD show the reconstruction performance of the rhythm representation, demonstrating that \M~with explicit rhythm modeling does the best job. In addition, the results indicate that RRD can be a valid metric for rhythm modeling. 

A visualization of output mel-spectrograms is shown in \autoref{fig:mels}. The effectiveness of rhythm modality modeling is clearly demonstrated: 1) \textit{\cite{parekh2020speech}} barely expresses the dynamic of the melody and the phonemes are scarcely distinguished; 2) the voiced parts are clustered temporally in \textit{SpeechSplit 2.0 (w/o SE)}, but the phonemes are still unrecognizable and the formants are disordered; and 3) \M~successfully re-align the phonemes in order. However, all of them reconstruct the pitch information in various degrees.

We set an additional baseline \textit{\M~(GAN)} by changing the diffusion decoder to a GAN-based decoder. Results indicate superior performance of diffusion models in singing voice synthesis. 

\subsection{Ablation Study}


\begin{table}[ht]
\centering
\setlength{\belowcaptionskip}{-0.4cm}
\begin{tabular*}{0.8\hsize}{l|cc}
\toprule
\bf Setting         & \bf CMOS-Q    & \bf CMOS-P  \\
\midrule
\textit{\M}         & 0.00          & 0.00        \\
\midrule[0.2pt]
\textit{w/o RA}     & -0.34         & -0.25       \\
\textit{w/o CM}     & -0.27         & -0.22       \\
\textit{w/o F0}     & -0.37         & -0.76       \\
\bottomrule
\end{tabular*}
\caption{\label{tab:ablation}
Ablation study results.
\textit{RA} denotes the rhythm adaptor, \textit{CM} denotes the cross-modal alignment, \textit{F0} denotes the final skip-connection of pitch representation.
}
\end{table}

As shown in \autoref{tab:ablation}, we conduct ablation studies to demonstrate the effectiveness of several designs in \M, including the rhythm adaptor, the cross-modal alignment, and the skip-connection of pitch representation. We conduct CMOS-Q (comparative mean opinion score of quality) and CMOS-P (comparative mean opinion score of prosody) evaluations on each setting: 1) \textit{w/o RA}: we stretch the speech rhythm representations defined in \autoref{eq:rhythm} using linear interpolation and use it for the subsequent cross-modal alignment. 2) \textit{w/o CM}: we drop the cross-modal alignment operation and stretch the content representation to the same length as the pitch contour and the adapted rhythm and simply fuse them together using element-wise addition. 3) \textit{w/o F0}: we cut off the skip-connection of pitch representation in the fusion, i.e., we only combine the adapted rhythm and the aligned content to the mel decoder. The results demonstrate a significant loss of performance when dropping any module. Specifically, it indicates that the rhythm adaptor plays an important role in modeling phonemes, replacing the rhythm adaptor will end up with a singing melody with unintelligible syllables. Removing the cross-modal alignment operation witnesses the degradation of both the audio quality and the prosody naturalness, but the presence of the adapted rhythm representation allows the mel decoder to implicitly re-align the content as thoroughly as possible, resulting in a slightly smaller quantity of loss compared to \textit{w/o RA}. As expected, The removal of F0 skip-connection drastically drops the prosody naturalness. 

\subsection{Zero-Shot Speech-to-Singing Conversion}

We conduct extensional experiments on zero-shot STS given only the singing samples and test the model on unseen speech samples. Specifically, we use the identical model architecture and the training pipeline, but carry out a singing-to-singing task and train the model to reconstruct singing samples from the proposed dataset. During the inference stage, we input the unseen speech signal and the target pitch contour to generate the corresponding singing samples. The results are also shown in \autoref{tab:main-results} and the task is denoted as \textit{\M~(zero-shot)}. \M~ demonstrates great potential in zero-shot STS. 


\section{Conclusion}
\label{sec:bibtex}

In this work, we presented \M, a speech-to-singing model based on modality disentanglement and cross-modal alignment. To achieve better voice quality and improve interpretability, we decomposed the input speech waveform into four variance information and proposed a novel cross-modal fusion mechanism. 
Specifically, we designed a rhythm adaptor to adjust the rhythm representation to deal with the altered pitch modality, and a cross-modal aligner to re-align the content representation. Finally, we conduct a cross-modal fusion to combine the different components together. 

Experimental results demonstrated that \M~achieved superior quality and naturalness compared with several baselines. Ablation studies demonstrated that each design in \M~was effective. Extensive experiments showed the great potential of \M~in zero-shot STS. We envisage that our work could serve as a basis for future STS studies.

\section{Limitations}
Research on the speech-to-singing conversion is important for human voice study and useful for practical applications such as computer-based music productions or entertainment. However, current STS approaches require an input condition of a fine-grained target F0 contour, which is always unavailable. In addition, the F0 contour of a singing utterance often possesses rich speaker-related information, which still needs further disentanglement. Finetuning F0 contours in real applications brings significant extra work. One of our future directions is to simplify the input conditions, such as musical scores. Furthermore, the preliminary attempt at the zero-shot STS task may lead to a better perspective.

Except for positive applications, STS systems may face ethical concerns. With the development of speech/singing voice synthesis technology, the cost of faking an utterance of a specific individual is gradually declining. Researchers need further consideration of the regulation and recognition of speech/singing voice synthesis.

\section{Acknowledgement}
This work was supported in part by the National Key R\&D Program of China under Grant No.2022ZD0162000, National Natural Science Foundation of China under Grant No.62222211, Grant No.61836002 and Grant No.62072397.

\bibliography{anthology,custom}
\bibliographystyle{acl_natbib}

\appendix

\begin{table*}[]
\setlength\tabcolsep{10pt}
\centering
\setlength{\belowcaptionskip}{-0.4cm}
\begin{tabular*}{0.8\hsize}{l|cc|cc}
\toprule
\bf Method                          & \bf LSD $\downarrow$  & \bf RCA $\uparrow$    & \bf MOS-Q $\uparrow$  & \bf MOS-P $\uparrow$  \\
\midrule
\textit{GT Mel}                     & 2.8974                & 0.9959                & 4.04$\pm$0.10         & 4.18$\pm$0.17         \\
\midrule[0.2pt]
\textit{SpeechSplit 2.0 (w/o SE)}   & 5.7681                & 0.9870                & 3.19$\pm$0.06         & 3.45$\pm$0.13         \\
\textit{SpeechSplit 2.0 (w/ SE)}    & \bf 4.4871            & 0.9848                & \bf 3.65$\pm$0.14     & \bf 3.88$\pm$0.05     \\
\midrule[0.2pt]
\textit{\M~(ours)}                  & 5.0129                & \bf 0.9934            & 3.58$\pm$0.19         & 3.81$\pm$0.09         \\
\bottomrule
\end{tabular*}
\caption{\label{tab:extensional}
Experimental results involving \textit{SpeechSplit 2.0 (w/ SE)}.
}
\end{table*}

\section{Architecture}
\label{appendix:architecture}

The architecture and hyperparameters are listed in \autoref{tab:arch}. We use three encoders E$_\text{P}$, E$_\text{R}$, and E$_\text{C}$ to encode the target F0, the target rhythm, and the source content features:
\begin{equation}
    \boldsymbol{x}_{\text{P}} = \text{E}_\text{P}(\boldsymbol{f}_0),\ \boldsymbol{x}_{\text{R}} = \text{E}_\text{R}(\boldsymbol{r}),\ \boldsymbol{x}_{\text{C}} = \text{E}_\text{C}(\boldsymbol{x})
\end{equation}
where all encoders are stacks of several convolution layers. 

These encoded features are fed into the rhythm adaptor, to both generate the discrete rhythm embeddings and train the rhythm predictor. The rhythm predictor consists of the cross-attention module and several convolutional layers, predicting the target discrete rhythm embeddings conditioned on the content and pitch features. The target rhythm embeddings are then used to re-align the source content representations. Finally, we carry out an inter-modality fusion across four different representations for re-synthesis.

Each encoder (E$_\text{R}$, E$_\text{P}$, and E$_\text{C}$) consists of two 1-D convolutional layers, where the kernel sizes are 7, 5, and 3, respectively. The linguistic hidden features extracted from wav2vec 2.0 are 32-dimensional, which are widely used for downstream tasks. The hidden size of all the model components is 256. The size of the codebook in the VQ module is set to 6. The cross-attention layer is implemented with one multi-head layer and a feed-forward layer, where the latter consists of a 1-D convolution layer and a fully-connected layer. 

\begin{table}
\centering
\setlength{\belowcaptionskip}{-0.4cm}
\begin{tabular*}{\hsize}{l|c|c}
\toprule
\multicolumn{2}{c|}{Hyperparameter}                            & \M    \\
\midrule
\multirow{3}*{\shortstack{Pitch\\Encoder}}              & Encoder Kernel            & 5     \\
~                                                       & Encoder Layers            & 3     \\
~                                                       & Encoder Hidden            & 256   \\
\midrule[0.2pt]
\multirow{3}*{\shortstack{Content\\Encoder}}            & Encoder Kernel            & 3     \\
~                                                       & Encoder Layers            & 2     \\
~                                                       & Encoder Hidden            & 256   \\
\midrule[0.2pt]
\multirow{3}*{\shortstack{Rhythm\\Encoder}}             & Encoder Kernel            & 7     \\
~                                                       & Encoder Layers            & 2     \\
~                                                       & Encoder Hidden            & 256   \\
\midrule[0.2pt]
\multirow{12}*{\shortstack{Rhythm\\Adaptor}}            & Attention Hidden          & 256     \\
~                                                       & Attention Heads           & 1     \\
~                                                       & Attention Layers          & 1   \\
~                                                       & Attention Window Size     & 0.5   \\
~                                                       & Attention Guided $g$ & 0.1        \\
~                                                       & Attention FFN Kernel      & 9     \\
~                                                       & Conv1D Kernel             & 3     \\
~                                                       & Conv1D Layers             & 2     \\
~                                                       & Conv1D Hidden             & 256   \\
~                                                       & Conv1D Dropout            & 0.8   \\
~                                                       & VQ Embeddings             & 6     \\
~                                                       & VQ Hidden                 & 256   \\
\midrule[0.2pt]
\multirow{7}*{\shortstack{Cross-\\Modal\\Aligner}}      & Attention Hidden          & 256     \\
~                                                       & Attention Heads           & 2     \\
~                                                       & Attention Layers          & 2   \\
~                                                       & Attention Dropout         & 0.1   \\
~                                                       & Attention Window Size     & 0.4   \\
~                                                       & Attention Guided $g$      & 0.1    \\
~                                                       & Attention FFN Kernel      & 9     \\
\midrule[0.2pt]
\multirow{4}*{\shortstack{Diffusion\\Decoder}}          & Denoiser Layers           & 20     \\
~                                                       & Denoiser Hidden           & 256     \\
~                                                       & Time Steps                & 4   \\
~                                                       & Noise Schedule Type       & VPSDE   \\
\midrule[0.2pt]
\multicolumn{2}{c|}{Total Number of Parameters}                     & 26M    \\
\bottomrule
\end{tabular*}
\caption{\label{tab:arch}
Hyperparameters of \M modules.
}
\end{table}

\begin{figure*}[ht]
\centering
\includegraphics[width=\textwidth]{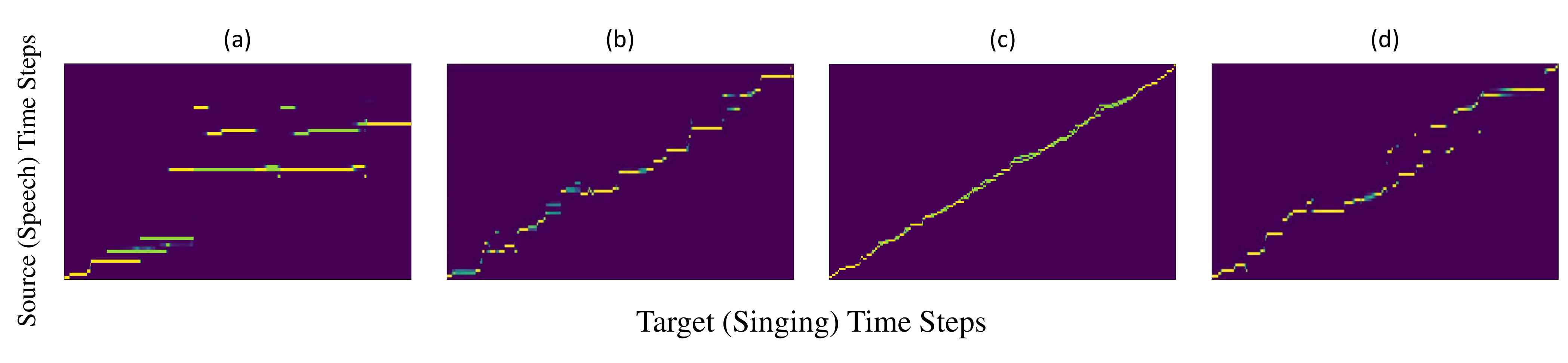}
\setlength{\abovecaptionskip}{-0.4cm}
\setlength{\belowcaptionskip}{-0.4cm}
\caption{Visualization of Attention weights. (a) Cross-attention without alignment regulation; (b) \textit{\M~(ours)}; (c) \textit{\M~(zero-shot)} during training stage; (d) \textit{\M~(zero-shot)} during inference stage.}
\label{fig:attention}
\end{figure*}

\section{Attention and Alignment Regulation}
\label{appendix:attention}

\subsection{Attention Windowing}
Attention windowing is a widely used technique that controls the "field of view" at each time step. Only a subsequence of key $\boldsymbol{\hat{x}} = [\boldsymbol{x}_{p_t-w},...,\boldsymbol{x}_{p_t+w}]$ are considered at each query time step $t$, where $w$ is the window width and $p_t$ is the middle position of the window along the key time axis. Specifically, we replace all the values outside the window with $-10^8$ before $\text{Softmax}(\cdot)$ so that the contribution outside the window is reduced significantly. 

\subsection{Guided Attention Loss}
To make sure that the attention weight matrix is nearly diagonal and monotonic, we adopt guided attention loss. Let $\alpha_{t,n}$ denotes the attention weight at query time step $t$ that attends to key time step $n$, a guided attention loss can be defined as:
    \begin{align}
        \mathcal{L}_{\text{attn}} &= \frac{1}{TN}\sum_{t=1}^T\sum_{n=1}^N \alpha_{t,n}w_{t,n}, \ \text{where} \\
                   w_{t,n} &= 1 - \exp\left(-\frac{\left(\frac{n}{N} - \frac{t}{T}\right)^2}{2g^2}\right)
    \end{align}
where T and N are the lengths of the query and the key, respectively. $w_{t,n}$ is the weight distribution of the constraint. $g$ is a hyperparameter used to control the concentration degree, which is set $g=0.1$ in practice. If $\alpha{t,n}$ is far from the diagonal, meaning that the key representations (like the content features) are placed in a random order, it provides a strong penalty. 


\subsection{Visualization}
A visualization of the attention weights is shown in \autoref{fig:attention}. All the attention weights are extracted from the last layer of the cross-modal aligner and averaged across attention heads. 

Comparing subfigure (a) the attention weights without alignment regulation and (b) that with regulation, the importance of attention path regulation is clearly demonstrated. Without alignment regulation, skips and non-monotonic situations occur, causing disordered or even indistinguishable phonemes. 

As for the zero-shot scenario, we compare the attention weights during the training stage and during the inference stage, which are shown in subfigures (c) and (d), respectively. Since the training is conducted in a self-supervised manner, the attention pattern demonstrates a perfectly linear pattern, as expected. The cross-modal aligner learns how to uniformly interpolate and stretch the input wav2vec 2.0 features to the same length as the target representations. However, in inference, the aligner is still capable of predicting the specific duration information of each linguistic unit of unseen speech data to a certain degree. As shown in subfigure (d), \M demonstrates its generalizability to unseen speech data and the ability to explore modality interaction in self-supervised pre-training.

\section{Dataset}
\label{appendix:dataset}

We collected and annotated the speech version of a subset of PopBuTFy to create a paired speech and singing dataset. During the collection, the private information of the speakers was protected. The qualified speakers are requested to read the lyrics of songs that have been sung by themselves. The personal vocal timbre is kept still during the recording process. We carefully select a subset of the collected recordings to create a high-quality dataset. In all, the dataset consists of 152 English pop songs ($\sim$5.5 hours in total) and the respective speech recordings ($\sim$3.7 hours in total) from 16 singers. All the audio files are recorded in a professional recording studio by professional singers, male and female. The recordings are sampled at 22050 Hz with 16-bit quantization. We randomly pick 111 pieces for validation and testing. 

\section{Extensional Experiments}
\label{appendix:extensional-experiments}

SpeechSplit 2.0 is originally designed for aspect-speciﬁc voice conversion, not STS tasks. Only manipulating the pitch component of SpeechSplit 2.0 input for STS may cause severe alignment problems, since the latent rhythm information is influenced. To lower the training difficulty and explore the importance of rhythm information, we add a new baseline \textit{SpeechSplit 2.0 (w/ SE)} that involves the target ground truth rhythm information, i.e., the perturbed spectral envelope, in the training and inference procedure. The performances are listed in \autoref{tab:extensional}. The results show a great improvement brought by this "information leak", in that the target rhythm information should not be available in a real situation. Also, the perturbed spectral envelope may still carry residual linguistic information for phoneme reconstruction.

\end{document}